\documentclass{aastex63}

\graphicspath{{./}{figures/}}

\begin{document}

\title{Low Dimensional Convolutional Neural Network For Solar Flares GOES Time Series Classification}

\author[0000-0003-3704-9003]{Vlad Landa}
\affiliation{ Department of Computer Science, Ariel University, Sciense Park, Ariel, 40700, Israel.}

\author[0000-0002-8902-5540]{Yuval Reuveni}
\email{yuvalr@ariel.ac.il}
\affiliation{Department of Physics, Ariel University, Ariel 40700, Israel.}
\affiliation{Eastern R\&D center, Ariel 40700, Israel.}
\affiliation{School of Sustainability, Interdisciplinary Center Herzliya, Herzliya 4610101, Israel.}



\begin{abstract}
Space weather phenomena such as solar flares, have massive destructive power when reaches certain amount of magnitude. Such high magnitude solar flare event can interfere space-earth radio communications and neutralize space-earth electronics equipment.
In the current study, we explorer the deep learning approach to build a solar flare forecasting model and examine its limitations along with the ability of features extraction, based on the available time-series data. For that purpose, we present a multi-layer 1D Convolutional Neural Network (CNN) to forecast solar flare events probability occurrence of M and X classes at 1,3,6,12,24,48,72,96 hours time frame. In order to train and evaluate the performance of the model,  we utilised the available Geostationary Operational Environmental Satellite (GOES) X-ray time series data, ranged between July 1998 and January 2019, covering almost entirely the solar cycles 23 and 24. The forecasting model were trained and evaluated in two different scenarios (1) random selection and (2) chronological selection, which were compare afterward. Moreover we compare our results to those considered as state-of-the-art flare forecasting models, both with similar approaches and different ones.The majority of the results indicates that (1) chronological selection obtain a degradation factor of 3\% versus the random selection for the M class model and elevation factor of 2\% for the X class model. (2) When consider utilizing only X-ray time-series data, the suggested model achieve high score results compare to other studies. (3) The suggested model combined with solely X-ray time-series fails to distinguish between M class magnitude and X class magnitude solar flare events. All source code are available at \href{https://github.com/vladlanda/Low-Dimensional-Convolutional-Neural-Network-For-Solar-Flares-GOES-Time-Series-Classification}{https://github.com/vladlanda}
\end{abstract}


\section{Introduction}\label{sec:intro}
A sudden burst of electromagnetic radiation originated at the Sun's surface, that travels with the speed of light and reach earth within 499.0 seconds, have the ability to interfere radio communication system, affect the Global Positioning System (GPS), neutralize space equipment, cause earth electric power blackouts and harm astronauts health, when reaches certain amount of magnitude. This electromagnetic burst known as "Solar Flares" phenomenon and can cause major blackout that easily exceed several billion dollars loss in repair and months of reconstruction, when reaches X class magnitude \citep{marusek2007solar,riswadkar2010solar}. Thus, building accurate and reliable solar flare forecast considering multiple range of time windows are essential for making decisions and taking protective manners in mission critical situations.

Attempts of constructing solar flares forecast takes place back in 1930s when \citep{giovanelli1939relations} suggested to examine the probability of an eruption based on the sunspot characteristics associated with it. Nowadays more and more studies leans toward Machine Learning (ML) algorithms that considered to be a data driven approaches. ML algorithms such as Support Vector Machine (SVM), Random Forest (RF) , Multi-Layer Preceptron (MLP) and Artificial Neural Network (ANN) where applied in the field of solar flares prediction. \citep{li2011solar} proposed an unsuperised clustering combined with vector quantity learning based on characteristics extracted form Solar and Heliospheric Observatory (SOHO)/Michelson Doppler Imager (MDI) data. \citep{yuan2010automated} had utilized photoshperic magnetic measurements for building automatic forecast of solar flares based on logistic regression and SVM model, to predict events that about to occur within the next 24 hours. Furthermore, \citep{huang2013improving} Developed forecasting model combining ${D}_{ARAL}$ (distance between active regions and predicted active longitudes) and parameters of the solar magnetic field based on instance-based learning model, while \citep{li2013solar} based a prediction solar flare model on MPL and learning vector quantization trained on a sequential sunspot data for 48 hours flare prediction. More recently \citep{muranushi2015ufcorin} introduced a fully automated solar flare prediction framework called UFCORIN (Universal Forecast Constructor by Optimized Regression of INputs) with two integrated regression models : SVM and handwriten linear regression algorithm, where \citep{nishizuka2017solar} examine and compare the performance of SVM, K-Nearest neighbors (k-NN) and extremely randomized trees performance on predicting maximum class of flares occurring in the next 24 hours while training the models on vector magnetograms, ultraviolet (UV) emission and soft X-ray emission available from Solar Dynamics Observatory (SDO) and GOES. \citep{bobra2015solar} study attempt to forecast M and X class magnitude solar flare event using SVM trained on SDO/Helioseismic and Magnetic Imager (HMI).

In the last decade, the advance of ML in the field of Deep Neural Networks, the massive growth of big-data and hardware advancement in Graphical Processing Units (GPUs) allowed for Artificial Intelligence (AI) algorithm to achieve human level performance in Computer Vision (CV) including image classification, object detection and segmentation \citep{lecun2015deep,russakovsky2015imagenet}. Moreover, AI have major successes in Natural Language Processing (NLP) that includes machine translation, image captioning and text generation \citep{wu2016google,vinyals2015show}. 

Recently, authors have applied Deep Neural Networks (DNN) for space weather predictions, in particular solar flares forecast. \citep{nagem2018deep} utilized the GOES X-ray flux 1-minute time series data for solar flare predictions by integrating three Neural Networks (NN) : first NN maps the GOES time series into Markov Transition Field (MTF) images. The second NN extract features form MTF. The third network are a Convolutional Neural Network (CNN) \citep{lecun1998gradient} that generates the prediction. Another study by \citep{chen2019identifying} propose to identify solar flare precursors by automated feature extractions and classify flare events from quite time for active regions as well as strong (X,M classes) versus weak (A,B,C classes) events at time frames of 1,3,6,12,24,48,72 hours before the event. Two types of models were examined : CNN and Recurrent Neural Network (RNN) based on Long Short Term Memory (LSTM) \citep{hochreiter1997long} cell trained on multiple data sources  : GOES, SDO/HMI. Furthermore, \citep{park2018application} presented a forecast application based on CNN model with binary outcome : 1 or 0 for daily flare occurrence of X,M and C classes. He compared his model to two well known models AlexNet \citep{krizhevsky2017imagenet} and GoogLeNet \citep{szegedy2015going} by training them utilizing transfer learning technique. Finally, \citep{huang2018deep} propose a deep learning method to learn forecasting patterns from line-of-sight magnetograms of solar active regions based on the available data from SOHO/MDI and SDO/HMI. His method forecast solar flare events at  6,12,24,48 hours window frame and compared to state-of-the-art forecasting models.

In this study, we propose a 1D CNN model designed as time-series classification for solar flare forecast application utilizing solely GOES-15 soft X-ray time-series data without hand-crafted features nor dedicated data preprocessing, compare to other studies. The suggested model take the GOES X-ray time-series data as an input and output a probability of flare occurrence of X class or M class. The current work, also examine the ability of such model design to learn and extract time-series features for distinguish between different solar flare class events.

The rest of the paper is organized in the following way : Data description and preparation are presented in Section \ref{sec:data}. The CNN architecture, training , evaluation processes and overall methodology are proposed in Section \ref{sec:method}. Model's performance and comparison are made in Section \ref{sec:results}. At last, the discussion and conclusions are provided in Section \ref{sec:discussion}.               
\section{Data} \label{sec:data}
We utilize the 1 minute average X-ray (0.1-0.8 nm) time-series data available from GOES \citep{schmit2013geostationary} mission  (\href{https://www.ngdc.noaa.gov/stp/satellite/goes/}{https://www.ngdc.noaa.gov/stp/satellite/goes/}). The first GOES (GOES-1) was launched in 1975 and operating by the United State's National Oceanic and Atmospheric Administration (NOAA)'s National Weather Satellite, Data, and Information Service division. All GOES mission spacecrafts is a geosynchronous satellite located at the height of around 35,800 km which provides full-disc view of the Earth as well as having unobstructed view of the sun. The main GOES mission are collecting infrared radiation and visible solar reflection from Erath's surface and atmosphere using Imager equipment as well as collecting atmospheric temperature, moisture profiles, surface and cloud top temperature, and ozone distribution using Sounder equipment. Moreover, GOES spacecraft carry on board the Space Environment Monitor (SEM) instrument consist of a magnetometer, X-ray sensor, high energy proton and alpha particle detector, and a energetic particles sensor. The X-ray sensor (XRS) found on board of GOES capable of registering two wavelength bands : 0.05-0.4nm and 0.1-0.8nm. Accourding to \href{https://www.spaceweatherlive.com/}{www.spaceweatherlive.com} the X-ray flux class defined by the magnitude of the long wave band (0.1-0.8nm) when reaches certain thresholds : ${10}^{-4}$,${10}^{-5}$ and ${10}^{-6}$ for X,M and C classes, respectively. GOES X-ray flux-data constitute the main source for confirming solar flare occurrence and used by previous studies to associated flare events in other data types \citep{chen2019identifying,huang2018deep}, meaning that GOES X-ray data source can be base for forecasting application without introducing any other sources of data. In order to form a sequential time-series X-ray data signal ranged from July 1998 to December 2019, multiple GOES mission sources are used, namely : GOES-10, GOES-14 and GOES-15. GOES-10 data is ranging from July 1998 to December 2009, GOES-14 data is ranging from January 2010 to December 2010 and GOES-15 data is ranging from January 2011 to December 2019. These three data sources are merged into one chronological sequence of 1 minute average X-ray signal, covering almost entirely solar cycle 23 and 24, from July 1998 to December 2009 and from January 2010 to December 2019, respectively. 
\subsection{Normalization, Scaling and Splitting}\label{sec2.1}
\begin{figure}[ht!]

\plotone{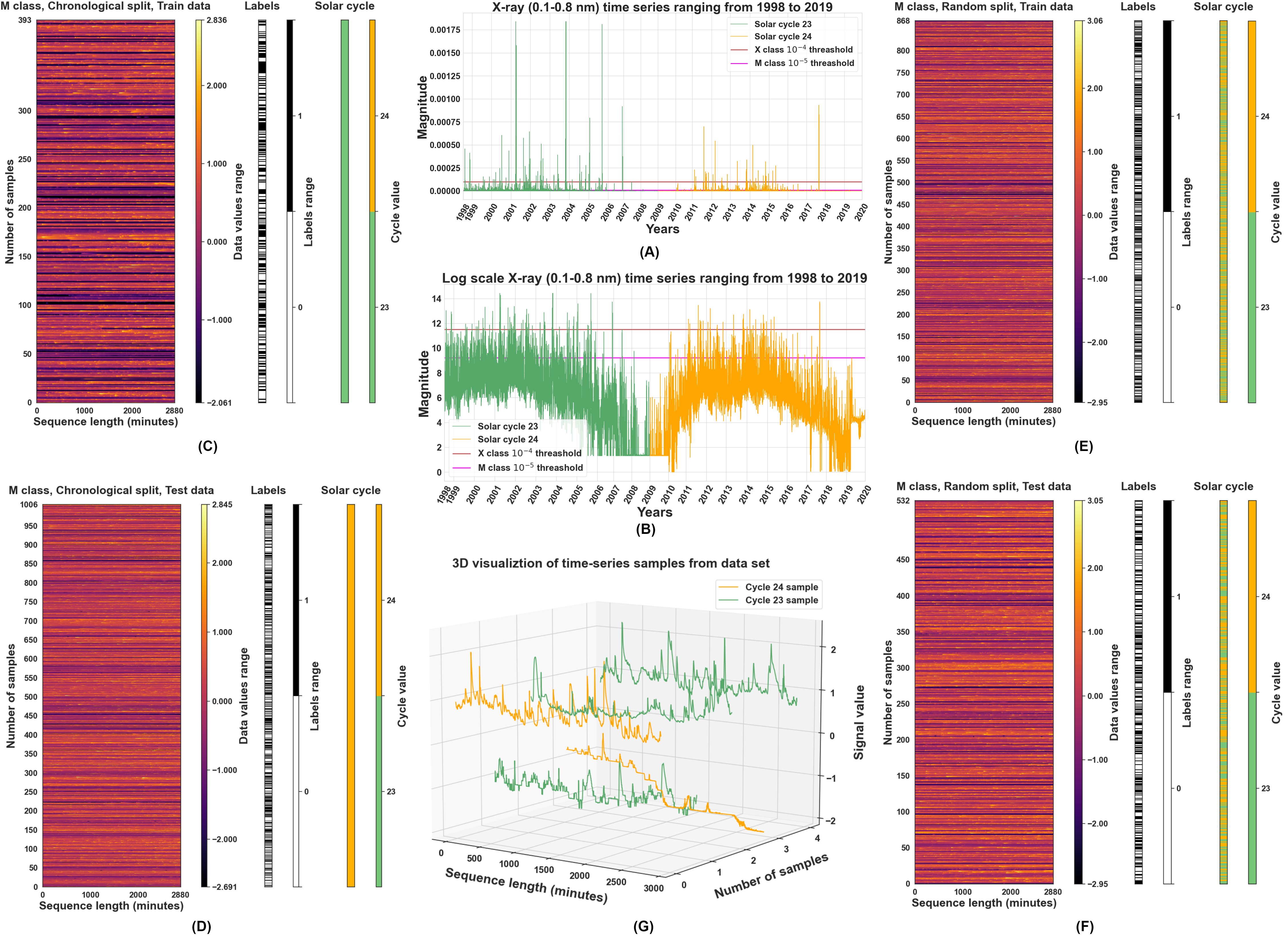}
\label{fig:data_pepare_vis}
\caption{(A) - Raw X-ray (0.1-0.8 nm) time series, with replaced minimum nominal value and scaled to 0 minimum value. (B) - X-ray time series log scaled. (C) - M class training data set chronological split, one hour ahead visualization. (D) - M class test data set chronological split,one hour ahead visualization. (E) - M class training data set random split,one hour ahead visualization. (F) - M class test data set random split,one hour ahead visualization. (G) - 3D visualization of 5 samples from random split test data set.}

\end{figure}

Utilizing the X-ray signal magnitude, we found all X and M solar flare events based on the corresponding thresholds associated with them : $1\cdot {10}^{-4}$ and $1\cdot {10}^{-5}$, respectively. In order to create two separate data sets for X and M solar flare classes with different prediction frames of 1,3,6,12,24,48,72 and 96 hours while preserving 48 hours of data as input to the model, we suggests the following scheme : first, we replace all missing values appears as "${-99999}$" in the time-series with GOES-15 nominal minimum ${1e}^{-9}$ value. That provides continues and smooth sequence, free of unexpected negative spikes, and it considered as ”None Event” or ”Quit Time” sequence. Then for every found solar flare event peak (M or X separately), confirm that higher magnitude event doesn't appears 12 hours ahead and no event appears of the same or higher magnitude 97 hours afore the peak (1 hour before the peak and 96 hours for prediction frames). Following that, a none event frame are selected by choosing random time point and confirming that no event higher than M class threshold appears 12 hours ahead or 97 hours afore. Moreover, the total variance of the selected frame do not appears to be below $1e^{-20}$ threshold, thus eliminating the use of frames with major nominal minimum value count. In such way an event/none event frame will have the length of 144 hours : 96 hours for prediction frames and 48 hours as input (for examination of 96 hours prediction). The total number of event frames for X class set and M class set counts : 171 and 1522 events, respectively, while none event frames set counts 1057 events. Each events frame set is split into training and testing sets by two different approaches : the Simple Random Sampling (SRS) approach and the chronological approach, see Figure \ref{fig:data_pepare_vis}. The SRS approach splits the set into training and testing by selecting samples with uniform distribution, i.e each sample in the set have equal probability to be selected as training or as testing. This approach is the most commonly used in most application and leads to low bias with balanced data \citep{reitermanovdata}. For the chronological approach we followed the suggested data split by \citep{park2018application} noting that data splitting methodology might influence model's forecasting performance and that random selection can increase model performance as training and testing events might be chosen from close time periods. Thus comparison of model's performance trained on same data but with different splitting technique are meaningful. Therefore, we selects data ranging from July 1998 to December 2009 as training and January 2010 to December 2019 as testing, in that case the training set consist of events only from solar cycle 23 and testing set consist of events only from solar cycle 24, making testing events to appears after training event chronologically. In addition every training and testing data set are structured to be with even flare/none flare numbers of events, based on the type event with smallest number of . Both splitting approaches are scale by ${1e}^{9}$ so its minimum value becomes 1.0. Afterwards, we applies the natural logarithm function upon the resulting sequence, such that the data maximum and minim range are narrower into range of $[0,log_{e}(1e^{-3}\cdot 1e^{9})]$, as the maximum nominal value of GOES-15 data set are $1e^{-3}$. Finally, we applies the standard normalization,which shifts the data to mean of 0 and scale the variance to be 1,on the training set and on the testing set based on the normalization parameters in the training sets.

\begin{deluxetable*}{c|ccc|ccc}
\tablenum{1}
\tablecaption{Number of X and M class events and None events per split type}
\tablewidth{0pt}
\tablehead{
    \colhead{} & \multicolumn{3}{c}{SRS} & \multicolumn{3}{c}{Chronological}\\
    \colhead{Sets} & \colhead{X class} & \colhead{M class} & \colhead{None Event} & \colhead{X class} & \colhead{M class} & \colhead{None Event}
}
\startdata
Train (70\%)                            & 84    & 434   & 84/434 & 87 & 197 & 87/197 \\
Test (30\% validation subtracted)       & 51    & 266   & 51/266 & 47 & 503 & 47/503 \\
Validation (30\% out of test)           & 36    & 186   & 36/186 & 37 & 84 & 37/84 \\
\hline
Total                                   & 171   & 886   & 1057   & 171 & 784 & 1057 \\
\enddata
\tablecomments{This table shows the number of available events of X class, M class and None events with SRS split approach based on 70\% for training and 30\% for testing, and the number of available events with Chronological split approach into solar cycle 23 and 24.}
\label{table:split_approach}
\end{deluxetable*}

\section{Method} \label{sec:method}
The CNN models has shown human level performance in the filed of computer vision and image processing, and currently is being deployed in self driving cars, flying drone systems, autonomous robotics, game playing and medicine. The core layers of such models are the convolutional layers. The convolutional layer consist of several filters also called kernals, which quantity and shape defined a priori to the training process with the hyper-parameters. When an input data(tensor) are passed into the convolutional layer, every kernal of that layer are convolved with the input tensor and generate a feature map, a general case of discrete 2-dimensional convolution is given by the following equation.
\begin{equation}
{f^{l}_{k}}_{map}[i,j] = \sum_{n = -N}^{N}\sum_{n = -M}^{M} x^{l-1}[i\cdot S +(N-P)-n,j\cdot S +(M-P)-m] \cdot w^{l}_{k}[n,m]
\end{equation}
Where ${f^{l}_{k}}_{map}[i,j]$ is the feature map $k$ of layer $l$ at index $i,j$, the $x^{l-1}$ is the output from previous layer $l-1$ that becomes the input to the current layer $l$ and $ w^{l}_{k}$ is the kernal of size $(2N+1) \times (2M+1)$. Two additional hyper-parameters are the stride $S$, which defines the kernal move step along the input tensor and the padding $P$ that pads the boundary of the input.

Due to the fact that convolution operation is linear and CNN is deep stack of linear layers combination, similar to DNNs, CNN is also designed with activation functions which allows modeling non-linear mapping from input domain into output domain. Often CNN models architectures includes the Rectified Linear Unit (ReLU) \citep{fukushima1982neocognitron,nair2010rectified}, that described as follows.

\begin{equation}
    ReLU(x) = max(0,x)
\end{equation}

\begin{figure}[ht!]
\plotone{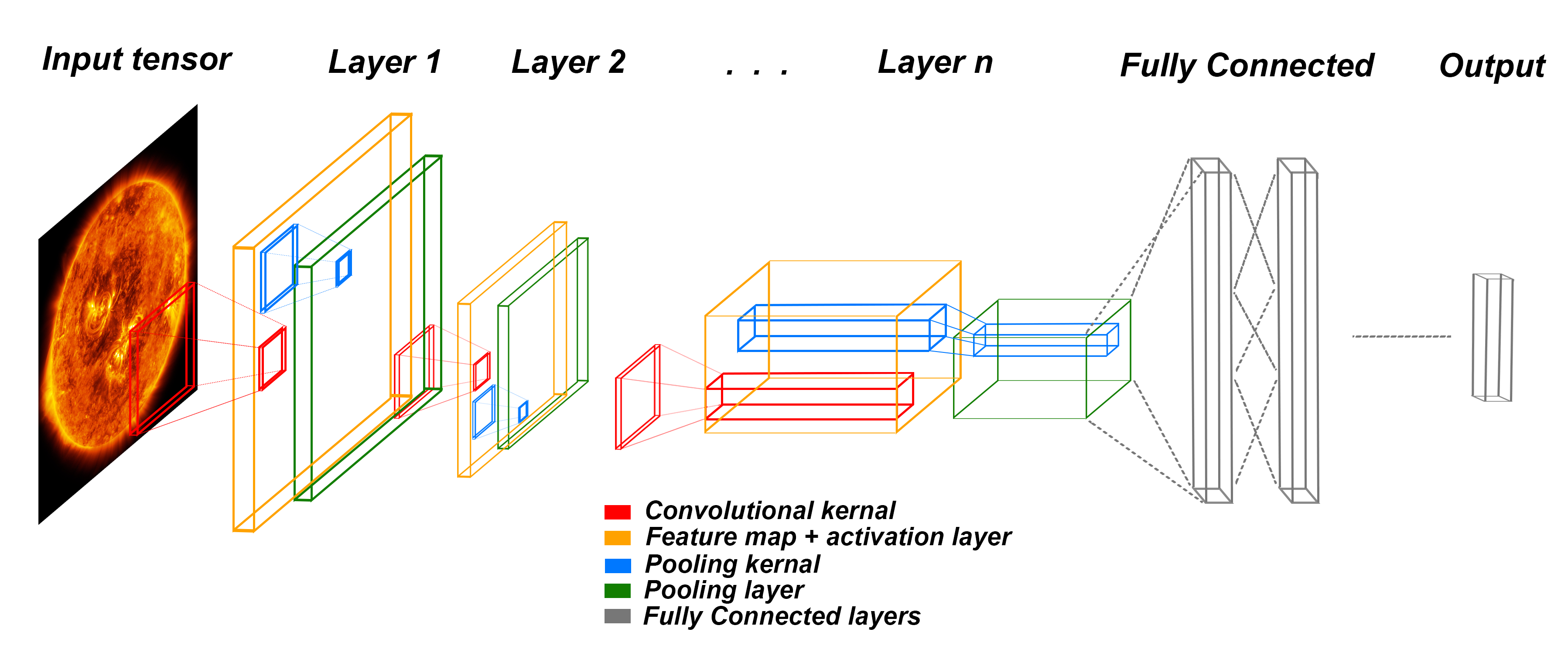}
\caption{General classification deep convolutional neural network stack of layers. Red - convolutional kernal. Orange - feature maps and activation function layer. Blue - pooling kernal. Green - pooling layers. Gray - fully conncected layers. The output tensor is passed though the softmax activation function and converted into a probability space.}
\label{fig:general_cnn_model}
\end{figure}

Furthermore, CNN feature maps encapsulate the spatial features found in the input tensor associated with the kernal's values that are learned during the training process. In general, input sample spatial features that describes the sample do not necessary grouped together in one location, but rather might be spread into different locations, therefore capturing those features leads to better model's performance. Thus, CNN models includes pooling layers, which pools information (based on the pooling layer type) from feature maps with applied activation function. Few pooling layer types are used in CNN models, one popular is the max pooling layer and defined by the following expression.
\begin{equation}
    {x^{l}_{k}}_{pool}[i,j] = \max_{\substack{-N' \leq n \leq N' \\ -M' \leq m \leq M'}}({f^{l}_{k}}_{map}[i \cdot S' + N' + n,j \cdot S' + M' + m])
\end{equation}
Where ${x^{l}_{k}}_{pool}[i,j]$ is the pooling tensor $k$ of layer $l$ at index $i,j$ operating on feature map ${f^{l}_{k}}_{map}$ with max pooling kernal  of size $(2N'+1) \times (2M'+1)$ and stride $S'$. $N',M',S'$ are defined by the hyper-parameters.
A general classification CNN model \citep{fawaz2019deep} consist of stacks of layers one after the other, such that the convolutional layer operates on the input tensor of the previous layer then it passed through the activation function, converting it into feature map and then the pooling layer pools spatial information about the feature. At the end of the model architecture a few fully connected layers followed by the softmax activation function, defined as:
\begin{equation}
    softmax({x})_{i} = \frac{\exp{x_{i}}}{\sum_{j=1}^{K} \exp{x_{j}}}
\end{equation}
Where ${x}$ is the input vector of real numbers and $K$ is the number of categories, maps the processed input into the output domain of categorical probabilities, see Figure \ref{fig:general_cnn_model}.

\subsection{Model architecture}
\begin{figure}[ht!]

\plotone{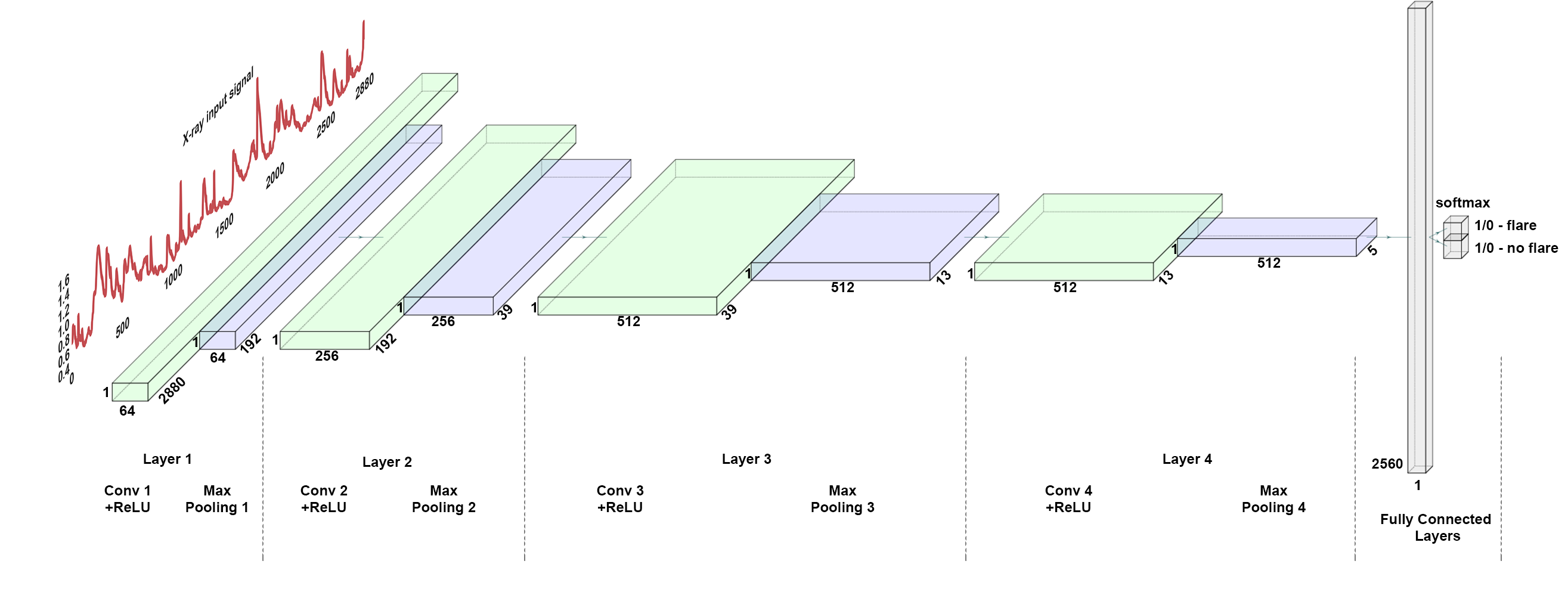}
\caption{Our convolutional neural network architecture with 4 convolutional layes with ReLU activation function, 4 max pooling layers and one fully connected layer with softmax activation function.}
\label{fig:1d_cnn_edited}
\end{figure}

We utilizing the general CNN architecture in order to develop a classification time series model as solar flare forecasting. In contrast to the general CNN model, that takes as an input two dimensional image, the X-ray time series data one-dimensional, hence we design an one-dimensional convolutional neural network based on the general CNN case \citep{wang2017time}. Our model, Figure \ref{fig:1d_cnn_edited}, consist of 4 convolutional layers, each layer followed by a ReLU activation function, 4 max pooling layers, a fully connected layer and an output layer with softmax activation function. In addition, every max pooling layer followed by a dropout layer, with dropout probability of 10\% (0.1 - hyper parameter value) for regularization and model overfitting avoidance \citep{srivastava2014dropout}.
\begin{itemize}
  \item The first convolutional layer ($conv1$) have 64 feature maps, kernal size of 30x30 and stride of 1, total $conv1$ size is 1x2880x64. The followed max pooling layer have kernals of size 15x15 with stride of 15 and a shape of 1x192x64.
  \item The second convolutional layer ($conv2$) have 256 feature maps, kernals with size 15x15 and a stride of 1, total $conv2$ size is 1x192x256, and its max pooling layer have kernals of size 5x5 with stride of 5 and a shape of 1x39x256.
  \item The third convolutional layer ($conv3$) have 512 feature maps, kernals with size 5x5 and a stride of 1, total $conv3$ size is 1x39x512. The followed max pooling layer have kernals of size 3x with stride of 3 and a shape of 1x13x512.
  \item The final convolutional layer ($conv4$) have 512 feature maps, kernals with size 3x3 and a stride of 1, total $conv4$ size is 1x13x512, and its max pooling layer have kernals of size 3x3 with stride of 3 and a shape of 1x5x512.
\end{itemize}
At the end of the model architecture a flattening layer flats the shape of the previous max pooling layer from 1x5x512 into 2560x1 connecting it to the output layer of size 2x1 making it a fully connected layer. The output layer passed through the softmax activation function to map the output into categorical probability space.       

\subsection{Data preparation \& Training}
\begin{figure}[ht!]

\plotone{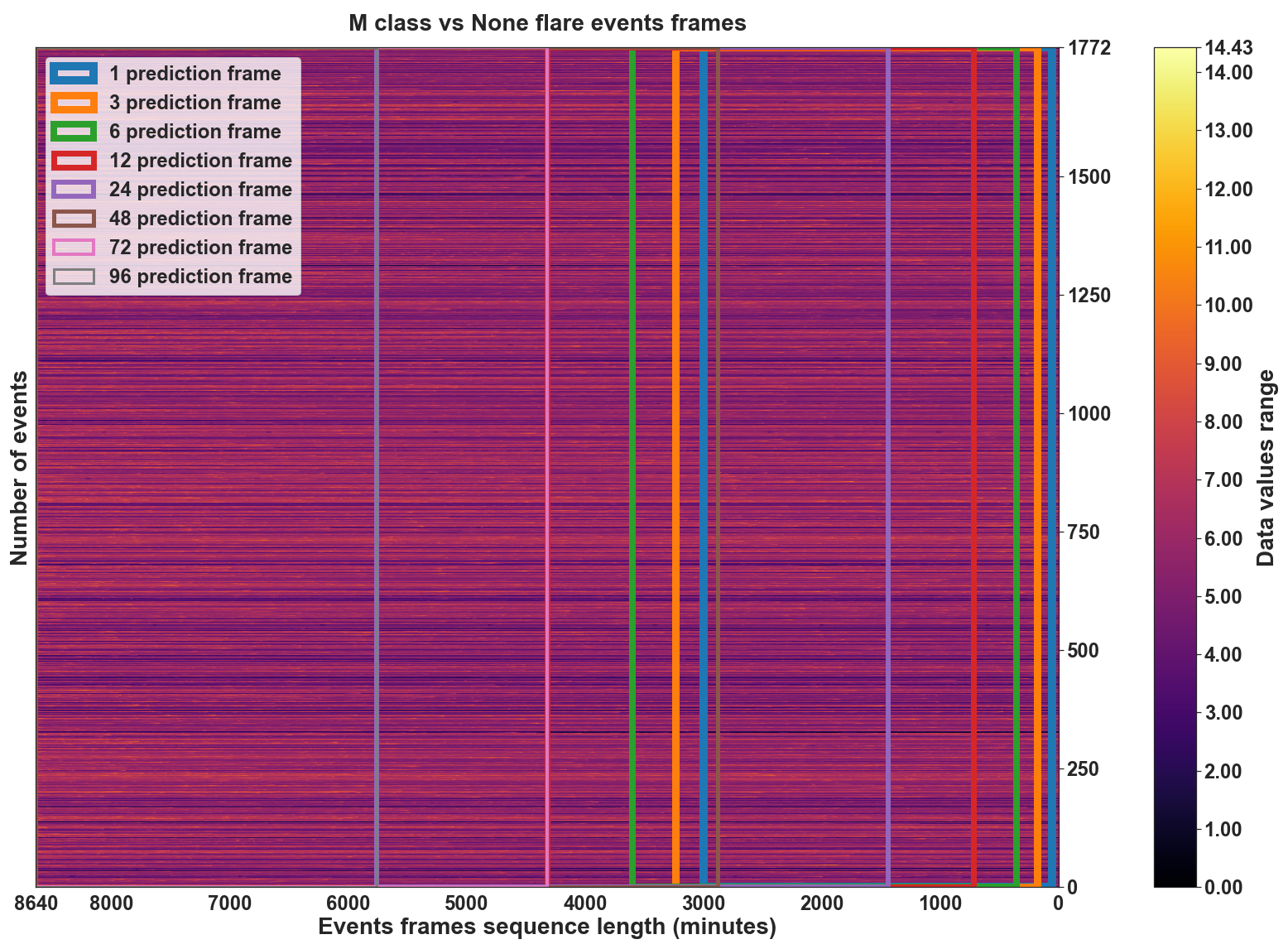}
\caption{Events frames of M class vs None class flares of random split data. The pulled 48 hours are color coded : blue - 1 hours, orange - 3 hours, green - 6 hours, red - 12 hours, purple - 24 hours, brown - 48 hours, pink - 72 hours, grey - 96 hours.}
\label{fig:pulling_48h}
\end{figure}

\begin{figure}[ht!]

\plotone{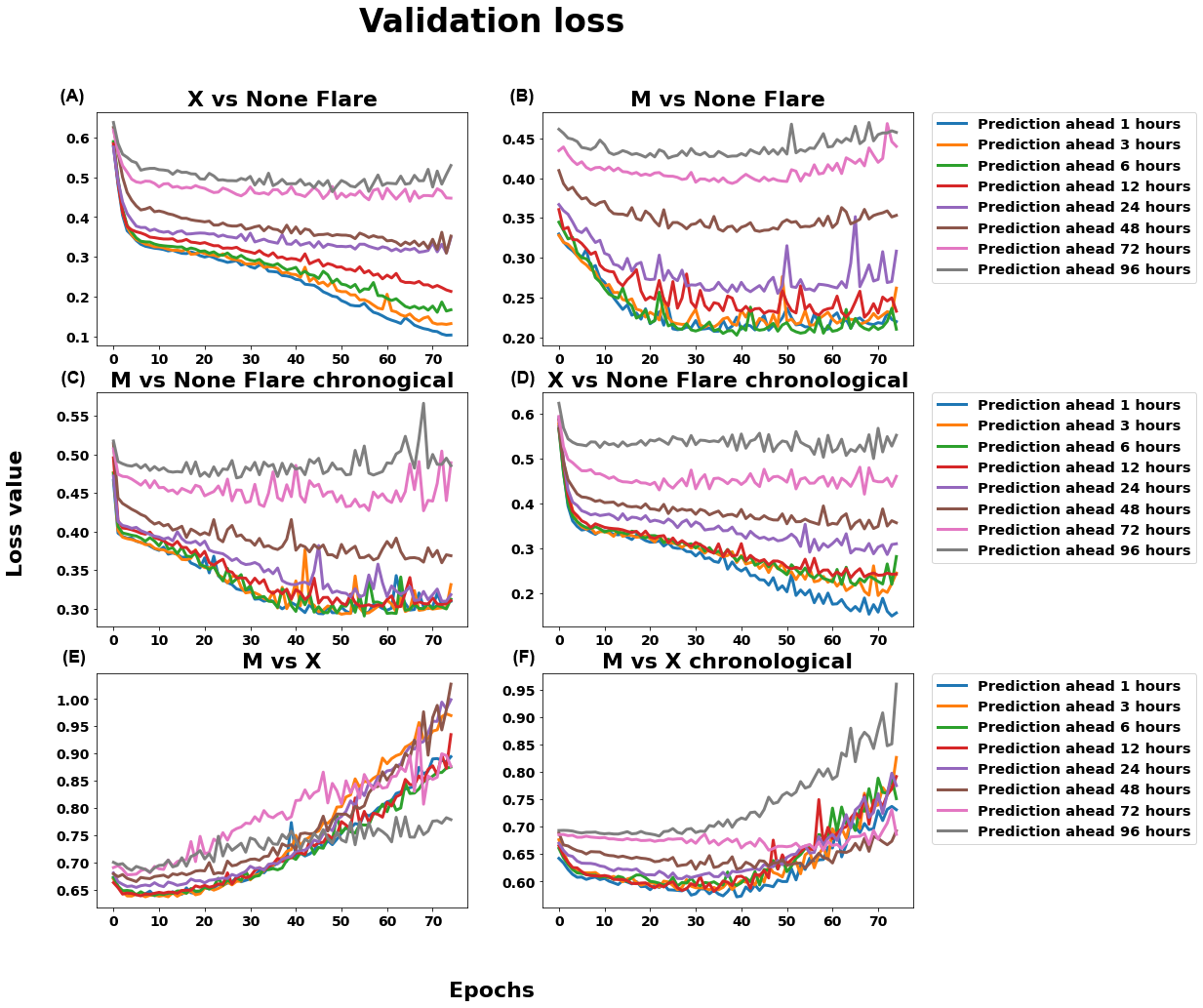}
\caption{Validation loss graphs of all 48 trained models. (A) - Validation loss of 8 different prediction window frames on X class flare vs None flare event with random data split. (B) - Validation loss of 8 different prediction window frames on M class flare vs None flare event with random data split. (C) - Validation loss of 8 different prediction window frames on M class flare vs None flare event with chronological data split. (D) - Validation loss of 8 different prediction window frames on X class flare vs None flare event with chronological data split. (E) - Validation loss of 8 different prediction window frames on M class flare vs X flare event with random data split. (F) - Validation loss of 8 different prediction window frames on M class flare vs X class flare event with chronological data split.}
\label{fig:validation_loss}
\end{figure}

In order to cover a wide range of prediction time frames, different data split methods and solar flare class types, we proposes to train individual model for each combination of the following categories.
\begin{enumerate}
    \item \label{itm:first} Solar flare class type : X-flare vs none-flare or M-flare vs none-flare (2 in total).
    \item \label{itm:second}Data split type : Chronological or Random (2 in total).
    \item \label{itm:third}Prediction time frame : 1,3,6,12,24,48,72,96 hours (8 in total).
\end{enumerate}
Ending up with 32 trained model, each for every combination (2x2x8 = 32). In addition, we explorer the ability of the proposed CNN architecture in distinguishing between M class solar flare and X class solar flare with every corresponding time frame of category \ref{itm:third} and every data split type of category \ref{itm:second}, leading to additional 16 trained models (2x8 = 16). In order to train the model in various prediction time frame, we pulled 48 hours window of data ,shifted by the prediction gap, out of the available 144 hours in the event frame, forming training and testing sets with 2880 minutes range, see Figure \ref{fig:pulling_48h}. In total we trained 48 individual models with the following hyper-parameters : we used the Adam \citep{kingma2014adam} optimizer with learning rate of $3\cdot 10 ^ {-5}$. We adopted  the Cross Entropy loss function for training. The Cross Entropy loss function is given by the following formula.
\begin{equation}
    CE(y,\hat{y}) = -\sum_{i=1}^{m}(y_ilog(\hat{y}_i) + (1-y_i)log(1-\hat{y}_i))
\end{equation}
Where $y$ is the ground truth one-hot encoded vector of size $m$, $\hat{y}$ is the model output prediction vector, also of size $m$, encoded with probabilities entries and sum up to 1. Further, we used a mini-batch size of 16 and 75 epochs in total. In addition an early stopping \citep{prechelt1998early} mechanism added to the training process for model's weights sharpshooting once the validation set loss gets to new minimum value, see Figure \ref{fig:validation_loss} for validation set loss graphs. 

\section{Results and Comparison} \label{sec:results}
Classifier's evaluation are build upon statistical scores it achieved on the test set. The scores are calculated according to $Confusion$ $Matrix$ \citep{fawcett2006introduction} that measures the performance of a machine learning algorithm based on four(4) different combinations of predicted and actual(ground truth) values, see Table \ref{table:confusion_matrix_table}. In this study we followed seven (7) common score metrics adopted by space weather community \citep{park2018application,bobra2015solar,huang2018deep,chen2019identifying}. 

\begin{deluxetable*}{c|cc}
\tablenum{2}
\tablecaption{Confusion matrix description table.}
\tablewidth{0pt}
\tablehead{
    \colhead{} & \colhead{Flare predicted} & \colhead{None-Flare predicted}
}
\startdata
Flare occurred (P)         & True Positive (TP)   & False Negative (FN) \\
Flare not occurred (N)    & False Positive (FP)   & True Negative (TN)  \\
\enddata
\tablecomments{ Confusion matrix description table. Flare occurred (P) - column of all the positive events. Flare not occurred (N) - column of all the negative events. True Positive(TP) - counts the numbers of positive events predicted by model as positive. False Negative(FS) - counts the numbers of positive events predicted by model as negative. False Positive(FP) - counts the numbers of negative events predicted by model as positive. True Negative (TN) - counts the numbers of negative events predicted by model as negative.}
\label{table:confusion_matrix_table}
\end{deluxetable*}

\begin{enumerate}
    \item Accuracy (ACC) - ration of the number of correct predictions, ranging form 0 to 1, while 1 is perfect accuracy score. Accuracy defined as follows. 
    \begin{equation}
        ACC = \frac{TP+TN}{P+N}
    \end{equation}
    \item Precision ( positive predicted value - PPV) - the ability not to label a negative event as positive, ranging from 0 to 1, while 1 is perfect precision score. Precision defined as follows.
    \begin{equation}
        PPV = \frac{TP}{TP+FP}
    \end{equation}
    \item Recall (true positive rate - TPR) - The ability to find all positive events, ranging from 0 to 1, while 1 is perfect recall score. Recall defined as follows.
    \begin{equation}
        TPR = \frac{TP}{TP+FN}
    \end{equation}
    \item F1-score (F1) - the ability of finding all positive events and not mis-classify negative. Measured by the harmonic mean of the precision and the recall, ranging between 0 and 1, while 1 indicating perfect precision and recall. F1 score defined by the following equation.
    \begin{equation}
        F1 = \frac{2\cdot PPV \cdot TPR}{PPV+TPR}
    \end{equation}
    \item Heidke Skill Score 1 (HSS$_1$) - ranging from -$\infty$ to 1, while 1 is perfect HSS$_1$ skill score. HSS$_1$ measures the improvement over a model that always predicts negative events (baseline model - HSS$_1$ = 0). HSS$_1$ defined as follows.
    \begin{equation}
        {HSS}_1 = \frac{TP+TN-N}{P} = \frac{TP-FP}{TP+FN} 
    \end{equation}
    \item Heidke Skill Score 2 (HSS$_2$) - ranging from -1 to 1, while 1 is perfect HSS$_2$ skill score. HSS$_2$ is a skill score compare to a random forecast. HSS$_2$ defined as follows.
    \begin{equation}
        {HSS}_2 = \frac{2\cdot TP\cdot TN - FN\cdot FP}{P\cdot(FN+TN)+N\cdot(TP+FP)} 
    \end{equation}
    \item True Skill Score (TSS) - Measures the difference between true positive and false positive rates, ranging from -1 to 1. Evaluated as the maximum distance of ROC curve from diagonal line. TSS defined as follows.
    \begin{equation}
        TSS = \frac{TP}{TP+FN} - \frac{FP}{FP+TN}
    \end{equation}
\end{enumerate}

\begin{figure}[H]
\plotone{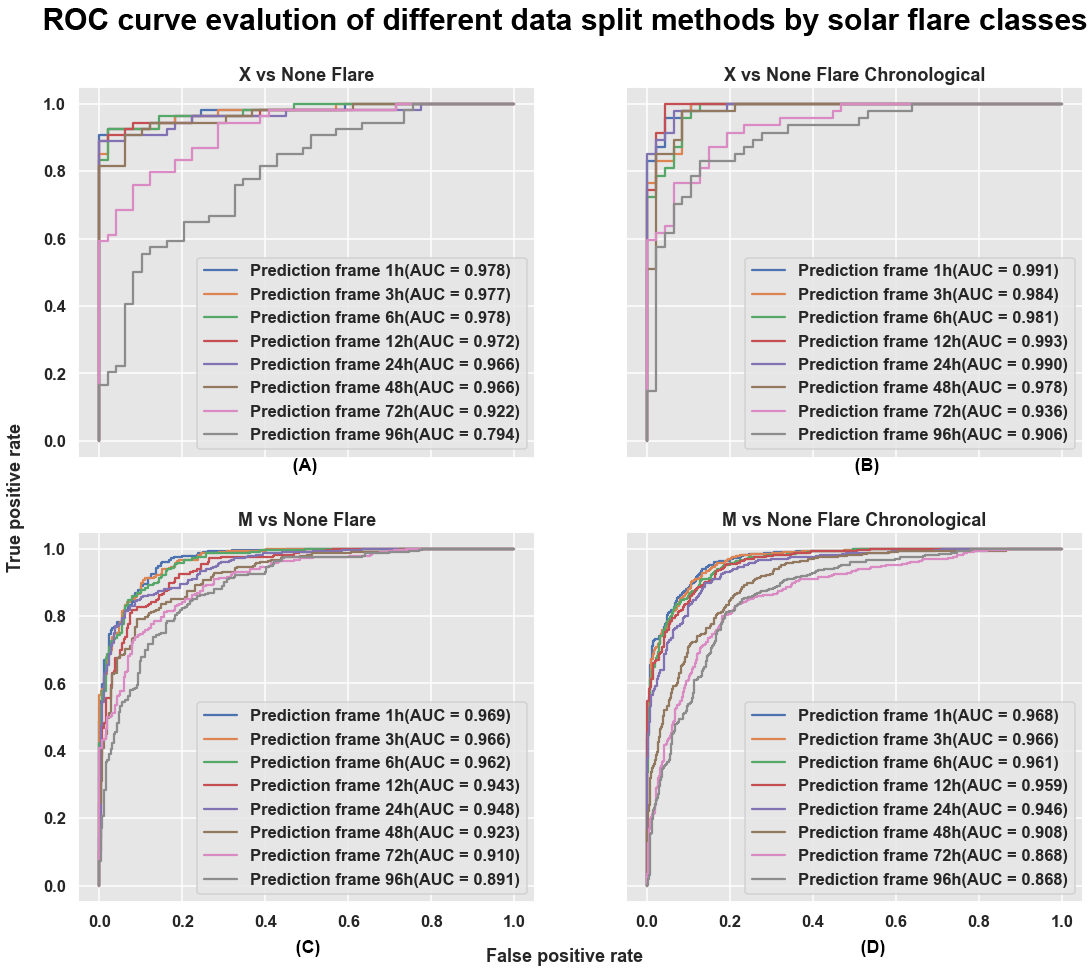}
\caption{ROC curve visualization of different splits types and different classification models by hours before the event. (A) - ROC curves of X class solar flare event vs none flare event classification on random split. (B) - ROC curves of X class solar flare event vs none flare event classification on chronological split. (C) - ROC curves of M class solar flare event vs none flare event classification on random split. (D) - ROC curves of M class solar flare event vs none flare event classification on chronological split.}
\label{fig:random_vs_chrono_roc}
\end{figure}

\begin{figure}[H]
\plotone{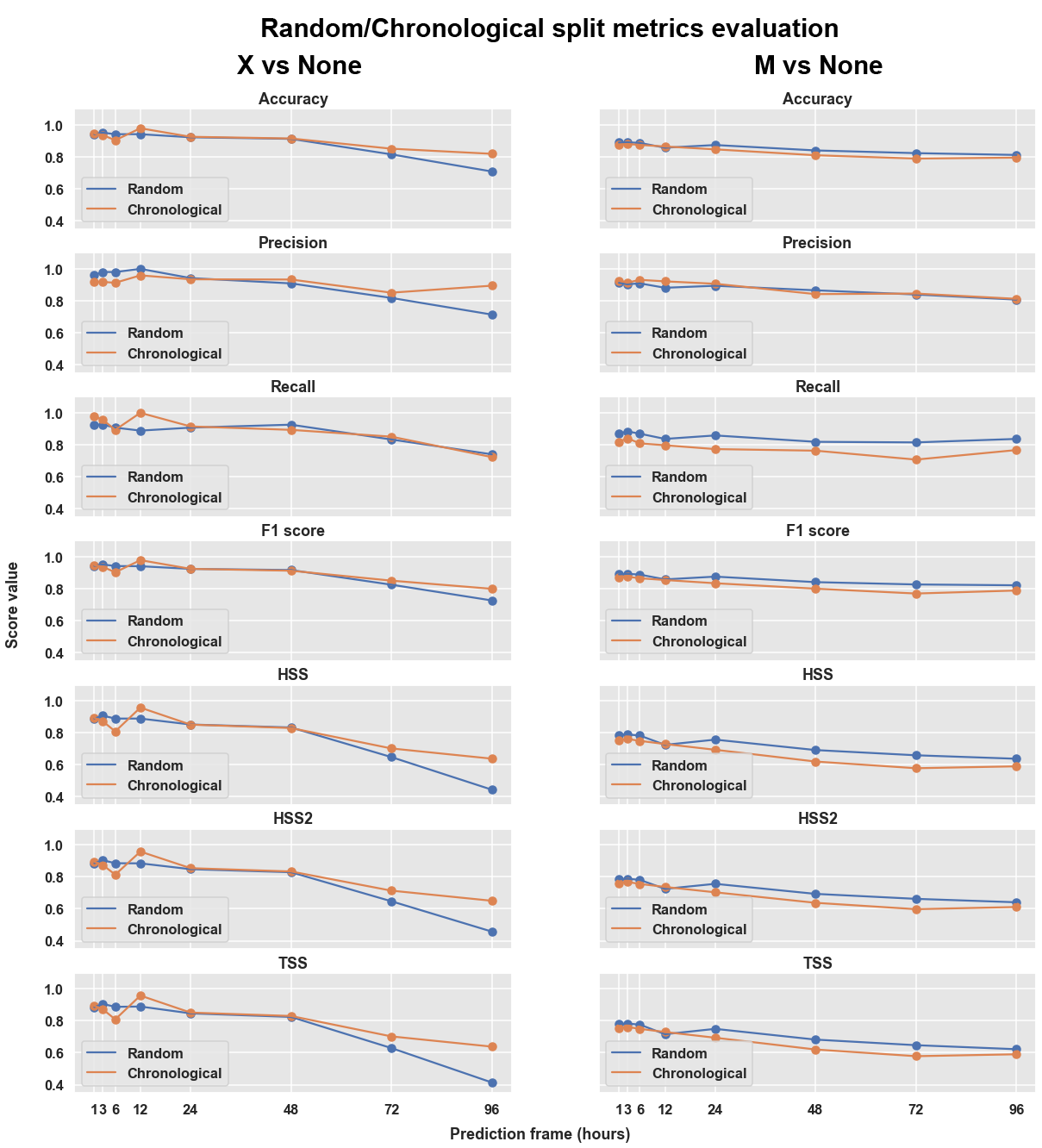}
\caption{ Visualization of random and chronological split types metrics evaluation. Left - metric evaluation of X class solar flare event vs none flare event classification model. Right -  metric evaluation of M class solar flare event vs none flare event classification model.}
\label{fig:random_vs_chrono_metrics}
\end{figure}

\begin{figure}[H]
\plotone{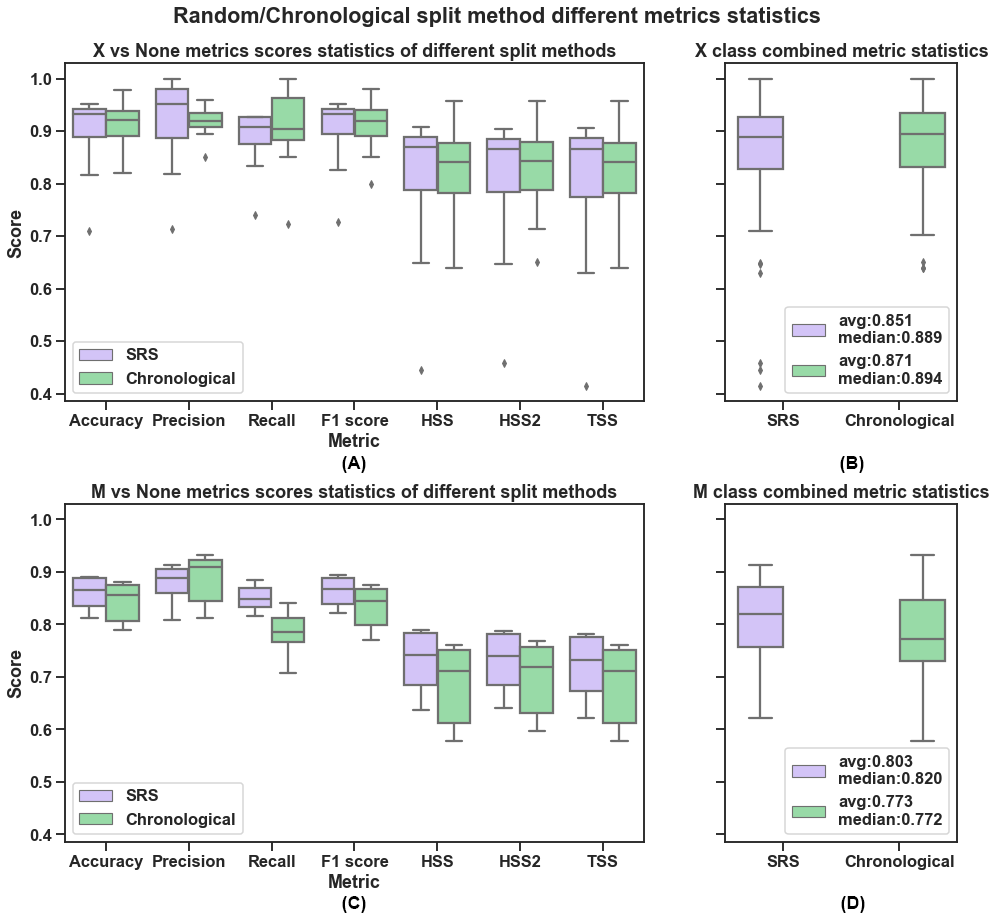}
\caption{Statistical visualization of different metrics by data split method. (A) - Metric statistics of X class solar flare event vs none flare event classification. (B) - X class solar flare classifier combined metrics statistics by split method. (C) - Metric statistics of M class solar flare event vs none flare event classification. (D) - M class solar flare classifier combined metrics statistics by split method.}
\label{fig:random_vs_chrono_statistics}
\end{figure}

\begin{figure}[H]
\plotone{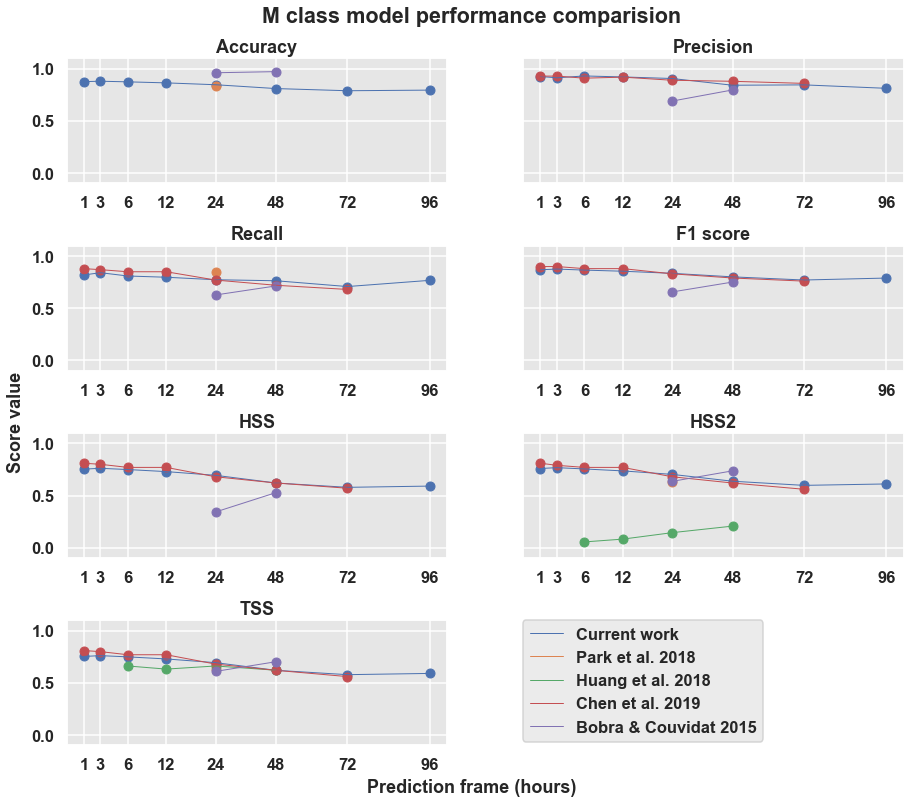}
\caption{M class solar flare classifier model metrics visualization comparison.}
\label{fig:m_class_model_performace_comparision}
\end{figure}

\begin{figure}[H]
\plotone{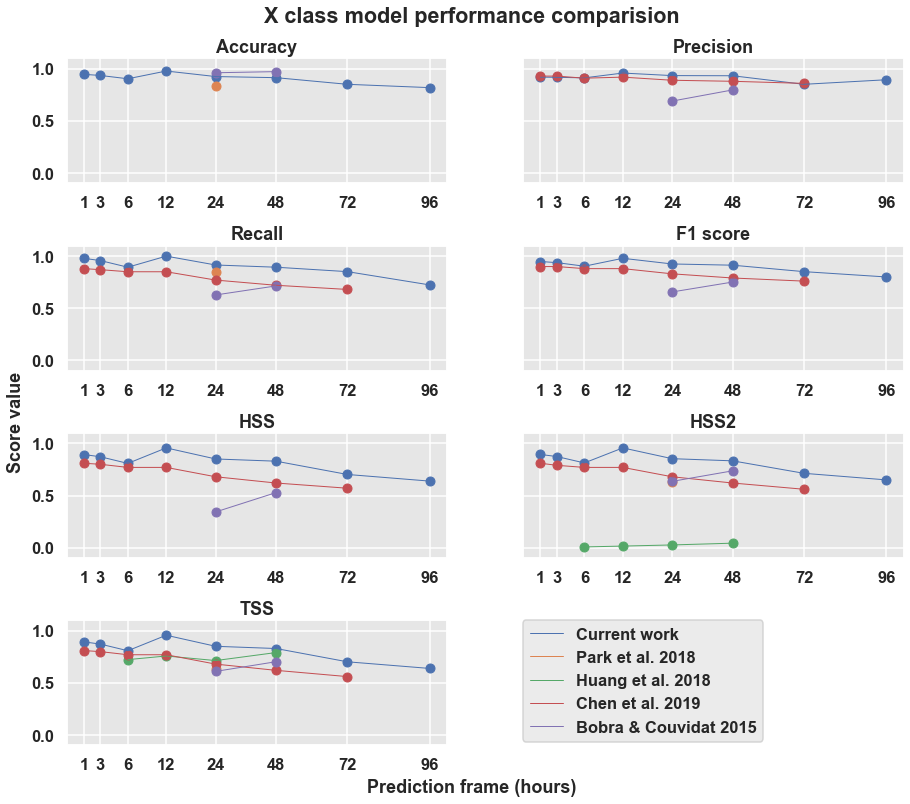}
\caption{X class solar flare classifier model metrics visualization comparison.}
\label{fig:x_class_model_performace_comparision}
\end{figure}

\begin{figure}[H]
\plotone{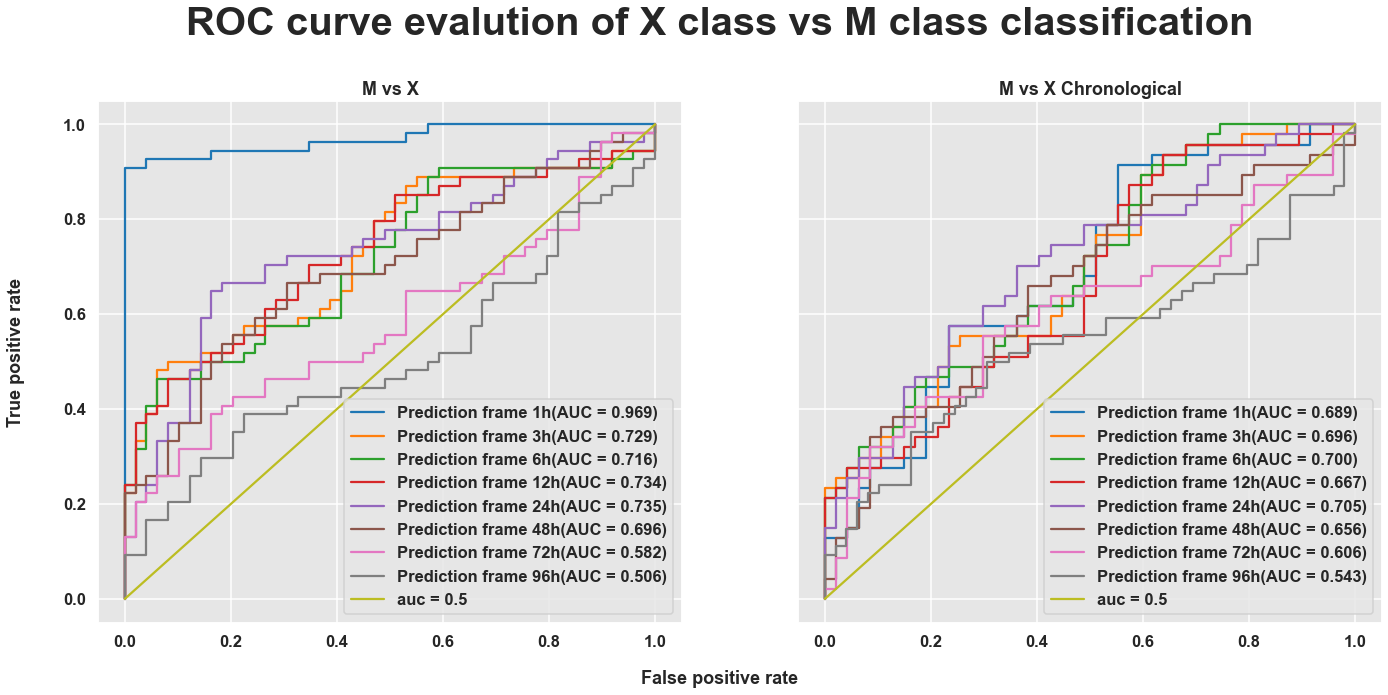}
\caption{ROC curve graph of model's distinguishing abilities between X and M solar flare classes on different prediction time frames . Left - random data split trained and tested model ROC curve. Right - chronological data split trained and tested model ROC curve.}
\label{fig:roc_curve_m_vs_x}
\end{figure}

As first evaluation we compare the results of two split method types : random and chronological, Figures \ref{fig:random_vs_chrono_roc} shows the Receiver Operating Characteristic (ROC) curve of each split method at every prediction time frame, separated by solar flare classes. In addition, Figure \ref{fig:random_vs_chrono_metrics} shows the comparison of different metrics skill scores of the two split types by the metrics, followed by a statistical analysis shown in Figure \ref{fig:random_vs_chrono_statistics}.Then, In order to create a compatible comparison with previous studies, we follow the data split methodology after \cite{park2018application}, who adopted the chronological split method as more suitable for space weather forecast systems. Therefore, current work comparison made utilizing solely the results achieved by training and testing the model on chronological data set split method. Four studies by \cite{chen2019identifying},\cite{park2018application},\cite{huang2018deep} and \cite{bobra2015solar} are chosen for comparison as the majority of them uses DNNs, except \cite{bobra2015solar} who uses SVM and they are considered as state-of-the-art flare forecasting models. All the comparisons made are based on the skill score metrics, Figures \ref{fig:m_class_model_performace_comparision} and \ref{fig:x_class_model_performace_comparision} shows visualization metrics of M class and X class solar flare classifier compare to previous works, respectively. An overall comparison of all prediction hours, metrics and models described in Table \ref{table:performance_comparison}. Moreover, the test evaluation of our model examining its ability to distinguish between whether M class or X class solar flare events is going to occur with both split method types are shown in Figure \ref{fig:roc_curve_m_vs_x} as ROC curves graphs.      

\begin{deluxetable*}{cc|cccccccc}
\tablenum{4}
\tablecaption{Performance comparison of X and M models train with chronological split.}
\tablewidth{0pt}
\tablehead{
    \multicolumn{2}{c}{} & \multicolumn{8}{c}{Number of hours ahead solar flare event} \\
    \cline{2-10} \colhead{Work} & \colhead{Metric} & \colhead{1hr} & \colhead{3hr} &\colhead{6hr} &\colhead{12hr} &\colhead{24hr} &\colhead{48hr} &\colhead{72hr} &\colhead{96hr} \\
}
\startdata
                                       & Accuracy     & 0.947& 0.936& 0.904& 0.979& 0.926& 0.915& 0.851& 0.819 \\
                                       & Precision    & 0.92& 0.918& 0.913& 0.959& 0.935& 0.933& 0.851& 0.895 \\
                                       & Recall       & 0.979& 0.957& 0.894& 1.0& 0.915& 0.894& 0.851& 0.723 \\
Current work                           & F1 score     & 0.948& 0.938& 0.903& 0.979& 0.925& 0.913& 0.851& 0.8 \\
  (X class)                            & HSS          & 0.894& 0.872& 0.809& 0.957& 0.851& 0.83& 0.702& 0.638 \\
                                       & HSS2         & 0.895& 0.874& 0.813& 0.957& 0.854& 0.833& 0.713& 0.65 \\
                                       & TSS          & 0.894& 0.872& 0.809& 0.957& 0.851& 0.83& 0.702& 0.638 \\
\hline
                                       & Accuracy     & 0.877& 0.881& 0.875& 0.865& 0.847& 0.81& 0.789& 0.795 \\
                                       & Precision    & 0.926& 0.914& 0.931& 0.922& 0.907& 0.842& 0.846& 0.813 \\
                                       & Recall       & 0.819& 0.841& 0.809& 0.797& 0.773& 0.763& 0.708& 0.767 \\
Current work                           & F1 score     & 0.869& 0.876& 0.866& 0.855& 0.835& 0.801& 0.771& 0.789 \\
  (M class)                            & HSS          & 0.753& 0.761& 0.75& 0.73& 0.694& 0.62& 0.579& 0.59 \\
                                       & HSS2         & 0.759& 0.768& 0.755& 0.736& 0.703& 0.637& 0.597& 0.611 \\
                                       & TSS          & 0.753& 0.761& 0.75& 0.73& 0.694& 0.62& 0.579& 0.59 \\
\hline
                                       & Precision    & 0.93& 0.93& 0.91& 0.92& 0.89& 0.88& 0.86 & ... \\
                                       & Recall       & 0.88& 0.87& 0.85& 0.85& 0.77& 0.72& 0.68 & ... \\
Chen et al. 2019                       & F1 score     & 0.9& 0.9& 0.88& 0.88& 0.83& 0.79& 0.76 & ... \\
                                       & HSS          & 0.81& 0.8& 0.77& 0.77& 0.68& 0.62& 0.57 & ... \\
                                       & HSS2         & 0.81& 0.79& 0.77& 0.77& 0.68& 0.62& 0.56 & ... \\
                                       & TSS          &0.81& 0.8& 0.77& 0.77& 0.68& 0.62& 0.56 & ... \\
\hline
                                       & Accuracy     & ...   & ... & ... & ... & 0.83 & ... & ... & ... \\
                                       & Recall       & ...   & ... & ... & ... & 0.85 & ... & ... & ... \\
Park et al. 2018                       & HSS2         & ...   & ... & ... & ... & 0.63 & ... & ... & ... \\
                                       & TSS          & ...   & ... & ... & ... & 0.63 & ... & ... & ... \\
\hline
Huang et al. 2018                      & HSS2         & ...   & ... & 0.054& 0.081& 0.143& 0.206 & ... & ... \\
                                       & TSS          & ...   & ... & 0.662& 0.632& 0.662& 0.621 & ... & ... \\
\hline
                                       & Accuracy     & ...& ... & ...& ... &0.962 & 0.973 & ... & ... \\
                                       & Precision    & ...& ... & ...& ... & 0.69 & 0.797 & ... & ... \\
                                       & Recall       & ...& ... & ...& ... &0.627 & 0.714 & ... & ... \\
Bobra and Couvidat 2015                & F1 score     & ...& ... & ...& ... &0.656 & 0.751 & ... & ... \\
                                       & HSS          & ...& ... & ...& ... &0.342 & 0.528 & ... & ... \\
                                       & HSS2         & ...& ... & ...& ... &0.636 & 0.737 & ... & ... \\
                                       & TSS          & ...& ... & ...& ... &0.61  & 0.703 & ... & ... \\
\enddata
\tablecomments{Full table of metrics comparison divided by prediction hours and skill scores.}
\label{table:performance_comparison}
\end{deluxetable*}

\section{Discussion and Conclusions} \label{sec:discussion}
In this study we design one dimensional convolutional neural network for time series classification as space weather forecast system. The network was trained solely on GOES X-ray time series data available at solar cycle 23 and 24. We focus on training two models, one model predicts X class solar flare events and second model predicts M class solar flares events. Both, X and M models trained for different prediction time frames afore the event and by two different data set split methods : random and chronological(trained on pass events and tested on future events). In both data split methods we kept the training and testing sets balanced. The evaluation of the models was done according to number of skill scores acceptable in space weather community for binary classification. For both models and split methods a degradation in performance grows as the prediction time frame increases, which is natural behaviour in forecast systems, the farther the forecast the more uncertainty presents \citep{camporeale2019challenge}. Despite \cite{nishizuka2017solar} notes regrading the influence of data sets separation on the forecast performance, in our case the performance difference between the chronological and random data set splits constitute 3\% for the M class model and only 2\% for the X class model in average over all the skill scores measured. Moreover, in the M class model case the difference in performance of 3\% express degradation in favor of random split, but in the X class model case the performance difference degradation is in favor of the chronological split. We chose to compare current work results achieved with models that was trained with chronological split method as suggested by \cite{park2018application}, the M class model results shows comparable high score to previous works. The work by \cite{chen2019identifying} shows higher results at a number of prediction time frames and few skill scores, but note that his work based on random split method. Other studies doesn't provide all the available skill score at the same prediction time frame as their focus is on different point of interest, but if considering the available skill score current work M class model achieves higher TSS values then those provided by \cite{park2018application,huang2018deep} at all prediction time frames and \cite{bobra2015solar} at 12 hour time frame but lower at 24 hour time frame. Same results are observable with the HSS2. \cite{park2018application} study achieve better performance with the recall score at 24 hour time frame and \cite{bobra2015solar} achieve higher accuracy score for 12 and 24 hours time frames. On the other hand, the X class model achieves higher performance over all compared studies and skill scores except to \cite{bobra2015solar} accuracy score values. Further, we examine the ability of the suggested model to distinguish between whether X class flare or M class flare are about to occur in different time frames. The results reveals that the model poorly capable of classifying between the two, such that at 96 hour time frame it achieve area under the curve (AUC of ROC curve) value of 0.506 with random split, which are equivalent to a model that flip a random binary coin. The topic of distinguishability between solar flare class require deeper research and are important in cases when a model trained for prediction of M or X class as one (binary predication whether X or M going to occur the model predicts 1 other wise 0) the recall skill score is calculated for both solar flare classes meaning it not truly express the false negative rate of X class flare the most harmful flare type.

\bibliography{main}{}
\bibliographystyle{aasjournal}



\end{document}